\documentclass{emulateapj}

\usepackage{float}
\usepackage{amsmath}
\usepackage{epsfig,floatflt}
\usepackage{subfigure}

\newcommand{\ce}{C_{\ell}^{\textrm{EE}}}
\newcommand{\cx}{C_{\ell}^{\textrm{TE}}}
\newcommand{\ct}{C_{\ell}^{\textrm{TT}}}
\newcommand{\se}{\sigma_{\ell}^{\textrm{EE}}}
\newcommand{\sx}{\sigma_{\ell}^{\textrm{TE}}}
\newcommand{\st}{\sigma_{\ell}^{\textrm{TT}}}
\newcommand{\sel}{\sigma_{\ell}}

\begin{document}

\title{Marginal distributions for cosmic variance limited CMB
  polarization data}

\author{H. K. Eriksen\altaffilmark{1,3,4} and I. K. Wehus\altaffilmark{2,5}}

\altaffiltext{1}{email: h.k.k.eriksen@astro.uio.no}
\altaffiltext{2}{email: i.k.wehus@fys.uio.no}

\altaffiltext{3}{Institute of Theoretical Astrophysics, University of
Oslo, P.O.\ Box 1029 Blindern, N-0315 Oslo, Norway}

\altaffiltext{4}{Centre of
Mathematics for Applications, University of Oslo, P.O.\ Box 1053
Blindern, N-0316 Oslo, Norway}

\altaffiltext{5}{Department of Physics, University of Oslo, P.O.\ Box
1048 Blindern, N-0316 Oslo, Norway}

\date{Received - / Accepted -}

\begin{abstract}
  We provide computationally convenient expressions for all marginal
  distributions of the polarization CMB power spectrum distribution
  $P(C_{\ell}|\sigma_{\ell})$, where $C_{\ell} =
  \{C_{\ell}^{\textrm{TT}}, C_{\ell}^{\textrm{TE}},
  C_{\ell}^{\textrm{EE}}, C_{\ell}^{\textrm{BB}}\}$ denotes the set of
  ensemble averaged polarization CMB power spectra, and $\sigma_{\ell}
  = \{\sigma_{\ell}^{\textrm{TT}}, \sigma_{\ell}^{\textrm{TE}},
  \sigma_{\ell}^{\textrm{EE}}, \sigma_{\ell}^{\textrm{BB}}\}$ the set
  of the realization specific polarization CMB power spectra. This
  distribution describes the CMB power spectrum posterior for cosmic
  variance limited data. The expressions derived here are general, and
  may be useful in a wide range of applications.  Two specific
  applications are described in this paper. First, we employ the
  derived distributions within the CMB Gibbs sampling framework, and
  demonstrate a new conditional CMB power spectrum sampling algorithm
  that allows for different binning schemes for each power spectrum.
  This is useful because most CMB experiments have very different
  signal-to-noise ratios for temperature and polarization. Second, we
  provide new Blackwell-Rao estimators for each of the marginal
  polarization distributions, which are relevant to power spectrum and
  likelihood estimation. Because these estimators represent marginals,
  they are not affected by the exponential behaviour of the
  corresponding joint expression, but converge quickly.
\end{abstract}


\maketitle

\section{Introduction}

During the last few decades cosmology has evolved from a data starved
branch of astrophysics, into a data driven high-precision science in
which theories may be subjected to stringent observational tests. This
revolution has to a large extent been driven by steadily improving
observations of the cosmic microwave background (CMB), allowing
cosmologists to have a close-up look at the very young universe. Two
leading experiments were the COBE-DMR \citep{smoot:1992} and WMAP
\citep{bennett:2003} satellite missions, while the third generation
experiment, Planck, will be launched late this year.

As observations continue to improve, increasingly demanding
requirements are imposed on the data analysis. While rather crude
approximations may be acceptable when interpreting low signal-to-noise
data, the situation is very different in the mid and high
signal-to-noise regime. Here, even ``small'' effects become clearly
visible, and may potentially compromise any cosmological
conclusion. Using accurate methods in this regime is critical. Some
real-world issues relevant to the CMB problem are non-cosmological
foregrounds, improper noise and/or beam characterization, and
sub-optimal likelihood approximations.

In 2004, a new approach to CMB analysis was proposed and implemented
by \citet{jewell:2004}, \citet{wandelt:2004} and \citet{eriksen:2004}.
Rather than taking the traditional approximate Monte Carlo approach
\citep[e.g.,][]{hivon:2002}, this new method employs the Gibbs
sampling algorithm to facilitate exact (in the maximum-likelihood
sense), global and efficient analysis of even high-resolution data
sets.  Equally important, the Gibbs sampling framework has unique
capabilities for error propagation, as it allows for easy
marginalization over virtually any auxiliary stochastic field. One
important example is that of non-cosmological foregrounds.

Since then, the method has been generalized to handle polarized CMB
data \citep{larson:2007} and joint foreground and CMB analysis
\citep{eriksen:2008a}, and has been applied most successfully to the
WMAP data \citep{odwyer:2004, eriksen:2007a, eriksen:2007b,
  eriksen:2008b}. Some useful examples of issues correctly identified
by the Gibbs sampler, but missed by other techniques, are 1) the
first-year WMAP likelihood bias at $\ell \lesssim 30$
\citep{eriksen:2007a, hinshaw:2007}, 2) foreground residuals in the
3-year WMAP polarization sky maps \citep{eriksen:2007b}, and 3)
residual monopole and dipole components in the 3-year temperature sky
maps \citep{eriksen:2008b, hinshaw:2008}. Following up on these
methodological advances, the WMAP team adopted the Gibbs sampler as a
central component in their analysis of the 5-year data, and, in fact,
their default low-$\ell$ temperature likelihood module is precisely
the Gibbs-based Blackwell-Rao code written and published by
\citet{chu:2005}.

While WMAP has done an excellent job on characterizing the large-scale
CMB temperature fluctuations, the current frontier in CMB science is
polarization. In just a few years, full-sky high-sensitivity data will
be available from Planck. And then, very likely, the situation will be
quite analogous to the one WMAP experienced in the temperature case:
Having robust, exact methods that allows for proper characterization
and propagation of systematics will be essential in the mid to high
signal-to-noise regime. The Gibbs sampler is among the leading
candidates to serve such a purpose.

Unfortunately, the Gibbs sampler, as currently described in the
literature, has two major limitations that needs to be resolved before
this promise can be fulfilled. First, the direct Gibbs sampler is
inherently inefficient in the low signal-to-noise regime, because the
step size between two consecutive samples is determined by cosmic
variance alone, whereas the full posterior width is determined by
noise.  Second, it is non-trivial to establish a full likelihood
approximation from the samples produced by the Gibbs sampler, because
of the dimensionality of the underlying space. Both of these issues
are currently under development, and reports are expected in the near
future (Jewell et al. 2008; Rudjord et al. and Eriksen et al., in
preparation).

In the present paper, we take a small but important first step towards
resolving these issues, by considering the marginal and conditional
densities of the probability distribution $P(C_{\ell}|\sigma_{\ell})$,
where $C_{\ell}$ is the ensemble averaged CMB power spectrum, and
$\sigma_{\ell}$ is the observed power spectrum of one given CMB
realization. This distribution plays a crucial role within the CMB
Gibbs sampling framework. On the one hand, it forms one of the two
conditionals in the main sampling scheme. On the other, it is the
kernel of the Blackwell-Rao estimator. Being able to describe this
analytically in different forms is therefore very useful. Two specific
applications are demonstrated in this paper, namely 1) a $C_{\ell}$
sampling algorithm that allows for different binning schemes in each
of the polarization components, and 2) Blackwell-Rao estimators for
each of $P(C_{\ell}^{TT}|C_{\ell}^{TE}, C_{\ell}^{EE},\mathbf{d})$,
$P(C_{\ell}^{TE}|C_{\ell}^{EE},\mathbf{d})$, and
$P(C_{\ell}^{EE}|\mathbf{d})$. Further applications will be
demonstrated in the papers mentioned above.

We also note that these expressions are completely general, and may
prove useful for any other method that considers both $C_{\ell}$ and
the CMB sky signal $\mathbf{s}$ as free variables. One such example is
the CMB Hamiltonian sampler recently developed by \citet{taylor:2007}.

\section{Notation and data model}
\label{sec:notation}

We now introduce a statistical model for the CMB observations,
and define our notation. First, we assume that the data may be
modelled by a signal and a noise term,
\begin{equation}
  \mathbf{d}(\hat{n}) = \mathbf{s}(\hat{n}) + \mathbf{n}(\hat{n}).
\end{equation}
Here, $\mathbf{d}$ is a 3-component $(T,Q,U)$ Stokes' parameter vector
observed in direction $\hat{n}$. $\mathbf{s}$ and $\mathbf{n}$ denote
similar vectors, describing the CMB field and instrumental noise,
respectively. Both the signal and the noise are assumed to be Gaussian
distributed with zero mean and covariances $\mathbf{C}$ and
$\mathbf{N}$, respectively.  (Note that we, for notational simplicity,
neglect real-world complications such as instrumental beams, frequency
dependent observations or foreground components in this expression;
the topic of this paper is the probability distribution $P(C_{\ell},
\textbf{s})$, and for this, all such issues are irrelevant.)

Next, we additionally assume the CMB field to be statistically
isotropic. It is therefore useful to decompose the $(T,Q,U)$ field
into spin-weighted spherical harmonics (see, e.g., Zaldarriaga and
Seljak 1997 for full details), with coefficients $(a_{\ell m}^T,
a_{\ell m}^E, a_{\ell m}^B)$.

Because the spherical harmonics are orthogonal on the full sky, and
$B$ has opposite parity of $T$ and $E$, the harmonic space CMB
covariance matrix is given by
\begin{align}
\mathbf{C}_{\ell m, \ell' m'} &= 
\left(\begin{array}{ccc} 
\left<a_{\ell m}^T a_{\ell' m'}^{T*}\right> & \left<a_{\ell m}^T
  a_{\ell' m'}^{E*}\right> & \left<a_{\ell m}^T a_{\ell'
    m'}^{B*}\right> \\
\left<a_{\ell m}^E a_{\ell' m'}^{T*}\right> & \left<a_{\ell m}^E
  a_{\ell' m'}^{E*}\right> & \left<a_{\ell m}^E a_{\ell'
    m'}^{B*}\right> \\
\left<a_{\ell m}^B a_{\ell' m'}^{T*}\right> & \left<a_{\ell m}^B
  a_{\ell' m'}^{E*}\right> & \left<a_{\ell m}^B a_{\ell'
    m'}^{B*}\right> 
\end{array}\right) \delta_{\ell\ell'} \delta_{mm'}\notag\\
&= 
\left(\begin{array}{ccc} 
C_{\ell}^{TT} & C_{\ell}^{TE} & 0 \\
C_{\ell}^{TE} & C_{\ell}^{EE} & 0 \\
0 & 0 & C_{\ell}^{BB}
\end{array}\right) \delta_{\ell\ell'} \delta_{mm'} \label{eq:Cl_covar}\\
&= C_{\ell} \,\delta_{\ell\ell'} \delta_{mm'}. \notag
\end{align}
In this expression, brackets denote ensemble averages, and the power
spectrum $C_{\ell}$ therefore corresponds to a theory spectrum,
similar to that produced by a Boltzmann code such as CMBFast
\citep{seljak:1996}. Note that $C_{\ell}$ denotes the matrix of all
power spectra, while a single component is indicated by subscripts
(e.g., $C_{\ell}^{TT}$).

One can also define the realization specific power spectrum,
$\sigma_{\ell}$, which is simply the averaged power in each multipole
for one given realization,
\begin{equation}
\sigma_{\ell} = \frac{1}{2\ell+1} \sum_{m=-\ell}^{\ell}
\left(\begin{array}{ccc} 
a_{\ell m}^T a_{\ell m}^{T*} & a_{\ell m}^T
  a_{\ell m}^{E*} & a_{\ell m}^T a_{\ell
    m}^{B*} \\
a_{\ell m}^E a_{\ell m}^{T*} & a_{\ell m}^E
  a_{\ell m}^{E*} & a_{\ell m}^E a_{\ell
    m}^{B*} \\
a_{\ell m}^B a_{\ell m}^{T*} & a_{\ell m}^B
  a_{\ell m}^{E*} & a_{\ell m}^B a_{\ell
    m}^{B*} 
\end{array}\right).
\label{eq:sigma_l}
\end{equation}
Explicitly, $C_{\ell}$ is the power spectrum corresponding to some
cosmological model, and $\sigma_{\ell}$ is the power spectrum of one
realization drawn from that model. It may therefore be useful to
imagine that observations of the CMB sky provide us with
$\sigma_{\ell}$, and from this we seek to constrain the underlying
cosmological theory, parametrized by $C_{\ell}$ and summarized by the
conditional distribution $P(C_{\ell}|\sigma_{\ell})$.

With this notation, it is straightforward to write down the joint
probability distribution for the CMB sky signal, $\mathbf{s}$, the CMB
power spectrum, $C_{\ell}$ and the data, $\mathbf{d}$,
\begin{align}
P(\mathbf{s}, C_{\ell}, \mathbf{d}) &= P(\mathbf{d}|\mathbf{s},
C_{\ell}) P(\mathbf{s}, C_{\ell})\notag\\
&\propto e^{-\frac{1}{2}(\mathbf{d}-\mathbf{s})^{\textrm{T}}
    \mathbf{N}^{-1}
    (\mathbf{d}-\mathbf{s})} P(\mathbf{s}, C_{\ell}).
\end{align}
and the CMB posterior distribution,
\begin{equation}
P(\mathbf{s}, C_{\ell}|\mathbf{d}) = \frac{P(\mathbf{s},
C_{\ell}, \mathbf{d})}{P(\mathbf{d})} \propto P(\mathbf{s},
C_{\ell}, \mathbf{d})
\end{equation}
These expressions involve two factors, namely the $\chi^2 =
(\mathbf{d}-\mathbf{s})^{\textrm{T}} \mathbf{N}^{-1}
(\mathbf{d}-\mathbf{s})$ and the CMB probability distribution
$P(\mathbf{s}, C_{\ell})$. Since we assume that the CMB field is
isotropic and Gaussian, as discussed above, the latter may be written
as (e.g., Larson et al.\ 2007)
\begin{align}
P(\mathbf{s}, C_{\ell}) &= P(\mathbf{s}|
C_{\ell}) P(C_{\ell})\notag\\
&\propto \frac{e^{-\frac{1}{2}\mathbf{s}^{\dagger}
    \mathbf{C}^{-1}
    \mathbf{s}}}{\sqrt{|\mathbf{C}|}} P(C_{\ell}) \notag\\
&= \prod_{\ell, m}\frac{e^{-\frac{1}{2}\mathbf{a}_{\ell m}^{\dagger}
    C^{-1}_{\ell} 
    \mathbf{a}_{\ell m}}}{\sqrt{|C_{\ell}|}} P(C_{\ell}) \notag\\
&=
\prod_{\ell}\frac{e^{-\frac{1}{2}\sum_{m}\textrm{tr}(\mathbf{a}_{\ell m}\mathbf{a}_{\ell m}^{\dagger}
    C^{-1}_{\ell})}}{|C_{\ell}|^{\frac{2\ell+1}{2}}}
P(C_{\ell}) \\
&=
\prod_{\ell}\frac{e^{-\frac{2\ell+1}{2}\textrm{tr}(\sigma_{\ell}C^{-1}_{\ell})}}
{|C_{\ell}|^{\frac{2\ell+1}{2}}} P(C_{\ell}) \notag\\
&=
\prod_{\ell} P(\sigma_{\ell}|C_{\ell}) P(C_{\ell}),\notag
\end{align}
where $P(C_{\ell})$ is a prior on
$C_{\ell}$. $P(\sigma_{\ell}|C_{\ell})$ is recognized as an inverse
Wishart distribution when interpreted as a function of $C_{\ell}$.

Before turning to the main topic of this paper, we recall that, for a
probability distribution $P(x,y)$, the marginal distribution is
defined by $P(x) = \int P(x,y) dy$, and the conditional by $P(y|x) =
P(x,y) / P(x)$. From these, one may also trivially derive Bayes'
theorem, $P(x|y) = P(y|x) P(x) / P(y)$. Therefore, for uniform priors
on both $C_{\ell}$ and $\mathbf{s}$, which we assume in this paper,
$P(\sigma_{\ell}|C_{\ell}) \propto P(C_{\ell}|\sigma_{\ell})$.

\section{CMB power spectrum distributions}
\label{sec:distributions}

\begin{deluxetable*}{lcl}
\tablewidth{0pt}
\tablecaption{CMB power spectrum distributions\label{tab:distributions}} 
\tablecomments{The determinant $\sel$
  denotes the two-dimensional $\{T,E\}$ determinant. See main text for
  definitions of 
  $I_1(\ell,A,B)$ and $I_2(\ell,C,D)$.} 
\tablecolumns{3}
\tablehead{Distribution & &Expression}
\startdata
\cutinhead{Joint $\{T,E\}$ distribution}
$P(\ct,\cx,\ce|\sigma_{\ell})$ & $\propto$ &
$\frac{|\sigma_{\ell}|^{\frac{2\ell-2}{2}}}{|C_{\ell }|^{\frac{2\ell+1}{2}}}
e^{-\frac{2\ell+1}{2}\textrm{tr}(\sigma_{\ell}C^{-1}_{\ell})}$ \\
\cutinhead{Bivariate marginals}
$P(\ct,\cx |\sigma_{\ell})$ & $\propto$ &
$\frac{|\sel|^{\frac{2\ell-2}{2}}(\ct)^{\frac{2\ell-3}{2}}}
{\big(\st(\cx)^2-2\sx\cx\ct+\se(\ct)^2\big)^{\frac{2\ell-1}{2}}}
e^{-\frac{2\ell+1}{2}\frac{\st}{\ct}}$ \\

$P(\cx,\ce |\sigma_{\ell})$ & $\propto$ &
$\frac{|\sel|^{\frac{2\ell-2}{2}}(\ce)^{\frac{2\ell-3}{2}}}
{\big(\se(\cx)^2-2\sx\cx\ce+\st(\ce)^2\big)^{\frac{2\ell-1}{2}}}
e^{-\frac{2\ell+1}{2}\frac{\se}{\ce}}$ \\

$P(\ct,\ce |\sigma_{\ell})$ & $\propto$ &
$\frac{{|\sel|}^{\frac{2\ell-2}{2}}}{(\ct\ce)^\ell}\cdot I_1(\ell,A,B)$\\

\cutinhead{Univariate marginals}
$P(C_{\ell}^{TT}|\sigma_{\ell})$ & $\propto$
&$\frac{(\sigma_{\ell}^{TT})^{\frac{2\ell-3}{2}}}{(C_{\ell}^{TT})^{\frac{2\ell-1}{2}}}
e^{-\frac{2\ell+1}{2} \frac{\sigma_{\ell}^{TT}}{C_{\ell}^{TT}}}$  \\ 
$P(C_{\ell}^{TE}|\sigma_{\ell})$ & $\propto$
&$\frac{{|\sel|}^{\frac{2\ell-2}{2}}}{(\st\se)^{\frac{2\ell-1}{2}}}
\cdot I_2(\ell,C,D)$  \\
$P(C_{\ell}^{EE}|\sigma_{\ell})$ & $\propto$ &
$\frac{(\sigma_{\ell}^{EE})^{\frac{2\ell-3}{2}}}{(C_{\ell}^{EE})^{\frac{2\ell-1}{2}}}
e^{-\frac{2\ell+1}{2} \frac{\sigma_{\ell}^{EE}}{C_{\ell}^{EE}}}$  \\ 
$P(C_{\ell}^{BB}|\sigma_{\ell})$ & $\propto$
&$\frac{(\sigma_{\ell}^{BB})^{\frac{2\ell-1}{2}}}{(C_{\ell}^{BB})^{\frac{2\ell+1}{2}}}
e^{-\frac{2\ell+1}{2} \frac{\sigma_{\ell}^{BB}}{C_{\ell}^{BB}}}$   \\

\cutinhead{Univariate marginals, one conditional variable} 

$P(C_{\ell}^{TT}|C_{\ell}^{TE}, \sigma_{\ell})$ &$\propto$ &
$\frac{(\st\se)^{\frac{2\ell-1}{2}}(\ct)^{\frac{2\ell-3}{2}}}
{\big(\st(\cx)^2-2\sx\cx\ct+\se(\ct)^2\big)^{\frac{2\ell-1}{2}} I_2(\ell,C,D)}
e^{-\frac{2\ell+1}{2}\frac{\st}{\ct}}$ \\

$P(C_{\ell}^{TE}|C_{\ell}^{TT}, \sigma_{\ell})$ &$\propto$ &
$\frac{{|\sel|}^{\frac{2\ell-2}{2}}(\ct)^{2\ell-2}}
{\big(\st(\cx)^2-2\sx\cx\ct+\se(\ct)^2\big)^{\frac{2\ell-1}{2}}
(\st)^{\frac{2\ell-3}{2}}}$ \\

$P(C_{\ell}^{TE}|C_{\ell}^{EE}, \sigma_{\ell})$ &$\propto$ &
$\frac{{|\sel|}^{\frac{2\ell-2}{2}}(\ce)^{2\ell-2}}
{\big(\se(\cx)^2-2\sx\cx\ce+\st(\ce)^2\big)^{\frac{2\ell-1}{2}}
(\se)^{\frac{2\ell-3}{2}}}$ \\

$P(C_{\ell}^{EE}|C_{\ell}^{TE}, \sigma_{\ell})$ &$\propto$ &
$\frac{(\st\se)^{\frac{2\ell-1}{2}}(\ce)^{\frac{2\ell-3}{2}} }
{\big(\se(\cx)^2-2\sx\cx\ce+\st(\ce)^2\big)^{\frac{2\ell-1}{2}} I_2(\ell,C,D)}
e^{-\frac{2\ell+1}{2}\frac{\se}{\ce}}$ \\

$P(\ct|\ce, \sigma_{\ell})$ &$\propto$ &
$\frac{{|\sel|}^{\frac{2\ell-2}{2}} I_1(\ell,A,B)}
{(\se)^{\frac{2\ell-3}{2}}(\ct)^l\sqrt{\ce}}
e^{\frac{2\ell+1}{2}\frac{\se}{\ce}}$\\

$P(\ce|\ct, \sigma_{\ell})$ &$\propto$ &
$\frac{{|\sel|}^{\frac{2\ell-2}{2}} I_1(\ell,A,B)}
{(\st)^{\frac{2\ell-3}{2}}(\ce)^l\sqrt{\ct}}
e^{\frac{2\ell+1}{2}\frac{\st}{\ct}}$
\enddata

\end{deluxetable*}

The main goal of this paper is to derive explicit expressions for the
marginals, and thereby the conditionals, of
$P(C_{\ell}|\sigma_{\ell})$. These are summarized in Table
\ref{tab:distributions}.

As seen above, $P(C_{\ell}|\sigma_{\ell})$ is given by the inverse
Wishart distribution, which, in $n$ dimensions and including the full
normalization factor \citep{gupta:2000}, reads
\begin{equation}
P(C_{\ell}|\sigma_{\ell}) =
\frac{(\frac{2\ell+1}{2})^{\frac{n(2\ell-n)}{2}}|\sigma_{\ell}|^{\frac{2\ell-n}{2}}}
{\Gamma_n(\frac{2\ell-n}{2})|C_{\ell
  }|^{\frac{2\ell+1}{2}}} 
e^{-\frac{2\ell+1}{2}\textrm{tr}(\sigma_{\ell}C^{-1}_{\ell})}.
\label{eq:joint_distribution}
\end{equation}
Here $\Gamma_n$ is the multivariate Gamma function. 

However, as discussed in Section \ref{sec:notation}, we are in this
paper interested in the special case for which $C_{\ell}^{TB} =
C_{\ell}^{EB} = 0$. In this case, the trace in equation
\ref{eq:joint_distribution} expands into a sum of two terms, and the
$C_{\ell}$ determinant factorizes into the product of a
two-dimensional $(T,E)$ determinant and $C_{\ell}^{BB}$. Thus, the
joint distribution factorizes as
\begin{equation}
  P(C_{\ell}|\sel) =
P(C_{\ell}^{TT}, C_{\ell}^{TE},
C_{\ell}^{EE}|\sel)P(C_{\ell}^{BB}|\sel).
\end{equation}
That is, $C_{\ell}^{BB}$ is independent of $(C_{\ell}^{TT},
C_{\ell}^{TE}, C_{\ell}^{EE})$, and follows a one-dimensional inverse
Wishart (or inverse Gamma) distribution. We will therefore not
consider the $BB$ component further in this paper. However, we note
that if one is interested in exotic models for which $\{TB, EB\}\ne
0$, the expressions derived in this paper will have to be revised
accordingly.

\subsection{The $(TE, EE)$ distribution, $P(C_{\ell}^{TE},
  C_{\ell}^{EE}|\sigma_{\ell})$}

We start by considering the two-dimensional marginal distribution
$P(C_{\ell}^{TE}, C_{\ell}^{EE}|\sigma_{\ell})$, which is obtained by
integrating $P(C_{\ell}^{TT}, C_{\ell}^{TE},
C_{\ell}^{EE}|\sigma_{\ell})$ over $C_{\ell}^{TT}$,
\begin{align}
&P(\cx, \ce| \sel)
=\int  P(\ct,\cx,\ce|\sigma_{\ell}) d\ct
\notag\\
&\propto |\sel|^{\frac{2\ell-2}{2}}\int_{\frac{(\cx)^2}{\ce}}^{\infty}
\left(\frac{1}{\ct\ce-(\cx)^2}\right)^{\frac{2\ell+1}{2}}
\\\notag
&\phantom{\propto |\sel|^{\frac{2\ell-2}{2}}\int^{\infty}}
\cdot e^{-\frac{2\ell+1}{2}\frac{\st\ce+\se\ct-2\sx\cx}{\ct\ce-(\cx)^2}}
d\ct.
\end{align}
Note that the lower limit in this integral is defined by $|C_{\ell}| > 0$,
since the power spectrum covariance matrix in equation
\ref{eq:Cl_covar} must be positive definite.

If we now define
\begin{equation}
k\equiv\se\frac{(\cx)^2}{\ce}-2\sx\cx+\st\ce,
\end{equation}
and make the change of variable
\begin{equation}
y=\frac{(2\ell+1)k}{2(\ct\ce-(\cx)^2)},
\end{equation}
this expression is transformed into
\begin{equation}
\begin{split}
P(\cx, \ce| \sel) \propto
\frac{|\sel|^{\frac{2\ell-2}{2}}}{\ce k^{\frac{2\ell-1}{2}}} 
 e^{-\frac{2\ell+1}{2}\frac{\se}{\ce}}
\int_0^\infty y^{\frac{2\ell-3}{2}}e^{-y} dy.
\end{split}
\end{equation}
The integral in this expression is simply the Gamma function,
\begin{align}
\Gamma\left(\frac{2\ell-1}{2}\right)=\int_0^\infty y^{\frac{2\ell-3}{2}}e^{-y} dy,
\end{align}
and, for our purposes, an irrelevant numerical normalization
factor. Thus, the final distribution reads
\begin{equation}
\begin{split}
  &P(C_{\ell}^{TE}, C_{\ell}^{EE}|\sigma_{\ell}) \propto
  \frac{|\sel|^{\frac{2\ell-2}{2}}}{\ce}
  e^{-\frac{2\ell+1}{2}\frac{\se}{\ce}} \cdot\\
  &\cdot\frac{1}
  {\left(\frac{\se{(\cx)^2}}{{\ce}}-2\sx\cx+\st\ce\right)^{\frac{2\ell-1}{2}}}
\end{split}
\label{eq:TE_EE}
\end{equation}
Note that because $TT$ and $EE$ occur symmetrically in equation
\ref{eq:joint_distribution}, the corresponding expression for
$P(\ct,\cx|\sel)$ is obtained simply by interchanging $EE$ and $TT$ in
equation \ref{eq:TE_EE}.

\subsection{The $(TT, EE)$ distribution, $P(C_{\ell}^{TT},
  C_{\ell}^{EE}|\sigma_{\ell})$}

Next, we consider $P(C_{\ell}^{TT}, C_{\ell}^{EE}|\sigma_{\ell})$,
which is obtained by integrating $P(C_{\ell}^{TT}, C_{\ell}^{TE},
C_{\ell}^{EE}|\sigma_{\ell})$ over $C_{\ell}^{TE}$. Unfortunately,
this distribution does not have a closed expression, but can instead
be written on the form
\begin{align}
P(\ct,\ce|\sel)\propto
\frac{{|\sel|}^{\frac{2\ell-2}{2}}}{(\ct\ce)^\ell}
\cdot I_{1}(\ell, A, B).
\label{eq:TT,EE}
\end{align}
Here $I_{1}$ denotes the integral
\begin{align}
I_1=\int_{-1}^{1}\frac{e^{-\frac{A-Bx}{1-x^2}}}
{(1-x^2)^{\frac{2\ell+1}{2}}}dx,
\end{align}
and we have defined the two dimensionless auxiliary parameters
\begin{align}
A&=\frac{2\ell+1}{2}\left(\frac{\st}{\ct}+\frac{\se}{\ce}\right) \\
B&=\frac{(2\ell+1)\sx}{\sqrt{\ct\ce}}
\end{align}
However, the fact that $I_{1}$ depends only on two dimensionless
parameters and $\ell$, implies that it can easily be
tabulated (and optionally splined for higher accuracy) for each
$\ell$, and thus computationally efficient lookup-tables may be
constructed. In most practical applications, which typically require
repeated evaluations of
$P(C_{\ell}^{TT},C_{\ell}^{EE}|\sigma_{\ell})$, Equation
\ref{eq:TT,EE} is therefore as useful for $(C_{\ell}^{TT},
C_{\ell}^{EE})$ as Equation \ref{eq:EE} is for
$(C_{\ell}^{TE},C_{\ell}^{EE})$, although implementationally a little
more complicated.

\subsection{The marginal $EE$ distribution,
  $P(C_{\ell}^{EE}|\sigma_{\ell})$}
\label{sec:EE}

We now compute the corresponding one-dimensional marginals, and begin
with $P(\ce|\sel)$, by integrating equation \ref{eq:TE_EE} over
$\cx$. This is simplified by introducing the new variable
\begin{equation}
y = \frac{\se\cx-\sx\ce}{\sqrt{|\sel|}\ce},
\end{equation}
The distribution then reads
\begin{align}
P(\ce|\sel)
&=\int P(\cx, \ce| \sel) d\cx
\notag\\
\propto \frac{(\se)^{\frac{2\ell-3}{2}}}
{\left(\ce\right)^{\frac{2\ell-1}{2}}}
&e^{-(\frac{2\ell+1}{2})\frac{\se}{\ce}}
\int_{-\infty}^\infty\left(\frac{1}{y^2+1}\right)^{\frac{2\ell-1}{2}}dy.
\end{align}
The integral in this expression is, for $\ell > 1$,
\begin{align}
  \int_{-\infty}^\infty\left(\frac{1}{y^2+1}\right)^{\frac{2\ell-1}{2}}dy
  =\frac{\Gamma(\frac{1}{2})\Gamma(\frac{2\ell-2}{2})}{\Gamma(\frac{2\ell-1}{2})},
\end{align}
which is a simple numerical constant. The desired marginal
distribution therefore reads
\begin{equation}
P(\ce|\sel) \propto \frac{(\se)^{\frac{2\ell-3}{2}}}
{\left(\ce\right)^{\frac{2\ell-1}{2}}}
e^{-(\frac{2\ell+1}{2})\frac{\se}{\ce}}.
\label{eq:EE}
\end{equation}
Again, we note that the corresponding expression for
$P(\ct|\sel)$ is obtained simply be replacing $EE$ with $TT$.

\subsection{The marginal $TE$ distribution,
  $P(C_{\ell}^{TE}|\sigma_{\ell})$}
\label{sec:TE}

Finally, we consider $P(C_{\ell}^{TE}|\sigma_{\ell})=\int_0^\infty
P(\cx,\ce|\sel)d\ce$.  As was the case for $P(C_{\ell}^{TT},
C_{\ell}^{EE}|\sigma_{\ell})$, this distribution does not have a
closed form, but may instead be written on a computationally
convenient form,
\begin{align}
P(\cx|\sel)\propto\frac{{|\sel|}^{\frac{2\ell-2}{2}}}{(\st\se)^{\frac{2\ell-1}{2}}}
\cdot I_{2}(\ell, C, D),
\label{eq:TE}
\end{align}
where $I_{2}$ denotes the integral
\begin{align}
I_2=\int_{0}^{\infty}\frac{x^{\frac{2\ell-3}{2}}e^{-\frac{1}{x}}}
{\left(\left(x-C\right)^2+D^2\right)^{\frac{2\ell-1}{2}}}dx.
\end{align}
The two dimensionless auxiliary parameters in this integral are
\begin{align}
C&=\frac{\sx\cx}{\st\se} \\
D&=\frac{\sqrt{|\sel|}\cx}{\frac{2\ell+1}{2}\st\se} 
\end{align}
Thus, as was the case for $I_1$, also $I_2$ may be tabulated over a
two-dimensional grid for each multipole. It is therefore
computationally straightforward to evaluate $P(\cx|\sel)$ at a given
value of $\cx$, even if it does not have a closed analytic expression.

\subsection{The conditional $TE$ distribution, $P(C_{\ell}^{TE} |
  C_{\ell}^{EE}, \sigma_{\ell})$}

From the above expressions, we may also derive all possible
conditional distribution, since $P(x|y) = P(x,y)/P(y)$. Here we will
only explicitly consider the conditional $TE$ distribution,
\begin{equation}
\begin{split}
&P(C_{\ell}^{TE}|C_{\ell}^{EE},\sigma_{\ell}) =
\frac{P(C_{\ell}^{TE}, C_{\ell}^{EE}
  |\sigma_{\ell})}{P(C_{\ell}^{EE}|\sigma_{\ell})} 
\\
&
\propto
\frac{(\ce)^{\frac{2\ell-3}{2}}}{(\frac{\sigma_{\ell}^{EE}
(C_{\ell}^{TE})^2}{C_{\ell}^{EE}} - 2\sigma_{\ell}^{TE}C_{\ell}^{TE} +
\sigma_{\ell}^{TT}C_{\ell}^{EE})^{\frac{2\ell-1}{2}}}
\frac{|\sigma_{\ell}|^{\frac{2\ell-2}{2}}}{(\sigma_{\ell}^{EE})^{\frac{2\ell-3}{2}}},
\end{split}
\end{equation}
which is relevant to several important applications (see, e.g.,
Section \ref{sec:applications}).

If we make the same linear transformation of $C_{\ell}^{TE}$ as in
Section \ref{sec:EE}, but including an additional $2\ell-2$ factor,
\begin{equation}
x = \frac{\sqrt{2\ell-2}}{\sqrt{|\sel|}\ce} \left(\se\cx-\sx\ce\right),
\end{equation}
this we see that this may be rewritten into a familiar form,
\begin{equation}
P(C_{\ell}^{TE}|C_{\ell}^{EE},\sigma_{\ell}) \propto
\frac{1}{(1+\frac{x^2}{2\ell-2})^{\frac{2\ell-1}{2}}}.
\end{equation}
We recognize this as the Student's t distribution with $\nu = 2\ell-2$
degrees of freedom. 

Since one of our goals is to sample from this distribution, it is
useful to have its cumulative distribution, $F$, on a closed form,
\begin{equation}
\begin{split}
F(C_{\ell}^{TE}|&C_{\ell}^{EE}, \sigma_{\ell}) =\\ &\frac{1}{2} +
x\Gamma(\frac{\nu+1}{2})
\frac{{_2}F_1(\frac{1}{2},\frac{\nu+1}{2};\frac{3}{2};-\frac{x^2}{\nu})}{\sqrt{\pi\nu}
\Gamma(\frac{\nu}{2})}.
\end{split}
\end{equation}
Here ${_2}F_1$ denotes the hypergeometric function (e.g., Abramowitz
\& Stegun 1972).

\section{CMB Gibbs sampling applications}
\label{sec:applications}

The analytic expressions derived in Section \ref{sec:distributions}
are the main results of this paper. Being completely general, these
may in principle be applied to a wide range of practical CMB
applications. However, they are particularly useful for methods that
consider both the sky signal $\mathbf{s}$ and the power spectrum
$C_{\ell}$ as free variables, such as the CMB Gibbs sampler. In this
section, we demonstrate two specific Gibbs-based applications, namely
1) a new $C_{\ell}$ sampling algorithm that supports general binning
schemes, and 2) new Blackwell-Rao estimators for marginal and
conditional distributions.

\subsection{A new $C_{\ell}$ sampling algorithm}
\label{sec:sampling}

\begin{figure}
\mbox{\epsfig{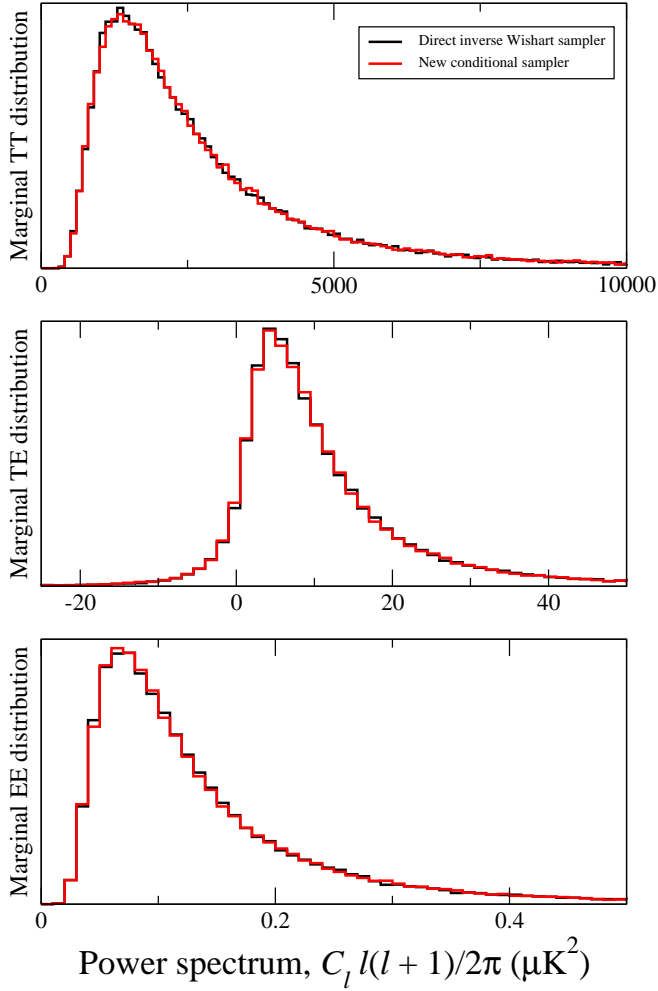}}
\caption{Comparison of marginal distributions obtained with the direct
  inverse Wishart sampler and with the new conditional sampler.}
\label{fig:samplers}
\end{figure}

The CMB Gibbs sampler draws samples from the joint posterior,
$P(\mathbf{s}, C_{\ell}|\mathbf{d})$, by alternately sampling from the
two corresponding conditionals (e.g., Jewell et al., Wandelt et al.,
Eriksen et al.\ 2004),
\begin{align}
\mathbf{s}^{i+1} &\leftarrow P(\mathbf{s}|C_{\ell}^{i}, \mathbf{d}), \\
C_{\ell}^{i+1} &\leftarrow P(C_{\ell}|\mathbf{s}^{i+1}).
\end{align}
(In this expressions, the arrow indicates sampling from the
distribution on the right hand side.) The former of these
distributions is a high-dimensional Gaussian distribution, while the
latter, which is the topic of the present paper, is a product of
independent inverse Wishart distributions. 

There is already a well known and simple algorithm available in the
literature to sample from the inverse Wishart distribution (e.g.,
Larson et al.\ 2007 or Gupta \& Nagar 2000): Let $p$ be the dimension
of the target matrix (e.g., $p=2$ for $\{T,E\}$), and $\Sigma_{\ell} =
(2\ell+1)\sigma_{\ell}$. Then the algorithms goes as follows: 1) Draw
$2\ell-p$ vectors from a Gaussian distribution with covariance matrix
$\Sigma_{\ell}$; 2) compute the sum of outer products of these
vectors; 3) invert this matrix. 

\begin{figure}
\mbox{\epsfig{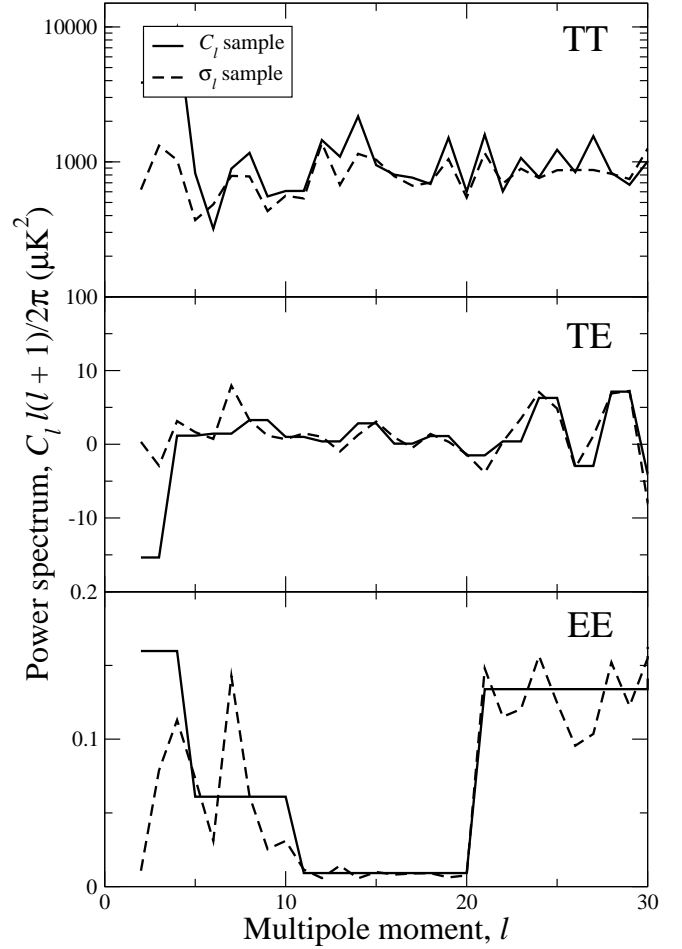}}
\caption{A single joint power spectrum sample $C_{\ell}$ drawn from
  $P(C_{\ell}|\sigma_{\ell})$, adopting individual binning schemes for
  $\ct$, $\cx$ and $\ce$.}
\label{fig:binning}
\end{figure}

This algorithm produce samples for a given multipole $\ell$. However,
in low signal-to-noise applications it is often desirable to bin many
multipoles together, in order to increase the effective
signal-to-noise of each variable. Because the CMB power spectrum
essentially scales as $\mathcal{O}(\ell^{-2})$, it is customary to bin
in units of $C_{\ell} \ell(\ell+1)/2\pi$. With this convention, the
above algorithm may be generalized to include binning by redefining
the covariance matrix as follows,
\begin{equation}
\Sigma_{\ell} = \sum_{\ell \in b} \frac{\ell(\ell+1)}{2\pi}
(2\ell+1)\sigma_{\ell}.
\end{equation}
Here $b = [\ell_{\textrm{min}},\ell_{\textrm{max}}]$ indicates the
multipole range of the bin under consideration. Note that the total
number of modes is now $M = (\ell_{\textrm{max}}+1)^2 -
\ell_{\textrm{min}}^2$, and therefore $M$ Gaussian vectors must be
drawn from $\Sigma_{\ell}$.

Unfortunately, this method has the serious drawback that the binning
scheme must be identical for $C_{\ell}^{TT}$, $C_{\ell}^{TE}$ and
$C_{\ell}^{EE}$. This is a problem because the signal-to-noise ratio
of most experiments is very different for $TT$ than for $EE$, and one
would lose much spectral resolution if one were to bin $C_{\ell}^{TT}$
with a bin size such that the signal-to-noise ratio for the
corresponding $C_{\ell}^{EE}$ bin is unity.

The new analytical expressions derived in Section
\ref{sec:distributions} allows us to resolve this problem. First, we
use the definition of a conditional distribution, and write 
\begin{equation}
\begin{split}
P(\ct, \cx, \ce|\sel) = &\,\,P(\ct|\cx,\ce,\sel) \,\cdot\\
&\cdot P(\cx|\ce,\sel) \,\cdot\\
&\cdot P(\ce|\sel).
\end{split}
\label{eq:factorization}
\end{equation}
The algorithm may now be written symbolically as follows,
\begin{align}
\bar{C}_{\ell}^{EE} &\leftarrow P(\ce|\sel) \\
\bar{C}_{\ell}^{TE} &\leftarrow P(\cx|\bar{C}_{\ell}^{EE}, \sel) \\
\bar{C}_{\ell}^{TT} &\leftarrow P(\ct|\bar{C}_{\ell}^{TE}, \bar{C}_{\ell}^{EE}, \sel).
\end{align}
Then $\{\bar{C}_{\ell}^{TT}, \bar{C}_{\ell}^{TE},
\bar{C}_{\ell}^{EE}\}$ will be a sample drawn from the joint
distribution $P(\ct, \cx, \ce|\sel)$. Note that each of these
conditional distributions is a one-dimensional distribution, and the
correlation with other polarization components comes into play only
conditionally, not jointly. One is therefore completely free to
specify the binning schemes for each component independently of the
others. 

We still need to write down a sampling algorithm for each of these
conditionals. However, since these are all one-dimensional, this is a
trivial task. The simplest approach, and the one currently implemented
in our codes, is to take advantage of the cumulative distribution:
Suppose we want to draw a sample from a univariate distribution,
$P(x)$, and have access to the corresponding cumulative distribution,
$F(x)$. Then we can draw a uniform variate $\eta \sim U[0,1]$, and
solve for $F(x) = \eta$. The sample $x$ will then be drawn from
$P(x)$.

However, a computational issue with this approach is the evaluation of
the cumulative distribution. With the notable exception of
$P(\cx|\ce,\sel)$ for a single multiple, we do not have explicit
analytic expressions for the cumulative distributions. Therefore, in
these cases we instead have to map out the analytic probability
densities over a grid, and do the integration
numerically. Fortunately, this requires only $\sim\mathcal{O}(10^2)$
function evaluations, and is therefore computationally quite fast. The
cost of the full Gibbs sampler is anyway hugely dominated by sampling
from the sky signal distribution $P(\mathbf{s}|C_{\ell}, \mathbf{d})$.

Nevertheless, one might want to consider alternative approaches for
applications in which this sampling step may be dominating. In such
cases, the rejection sampler (e.g., Liu 2001) appears as a promising
candidate. In this approach, one samples from an auxiliary
distribution that preferably should approximate the target
distribution quite well. Since our distributions are all one
dimensional, and strongly single-peaked, it should not be difficult to
establish such auxiliary functions. Indeed, the Student's t
distribution itself is a typical candidate for such purposes, since it
has a quite heavy tail.

In Figure \ref{fig:samplers} we compare the marginal distributions
obtained by the direct inverse Wishart distribution and by the new
sampler presented here, for $\ell=10$ only. As expected, they agree
perfectly. Next, in Figure \ref{fig:binning} we show a single sample
from the joint all-$\ell$ distribution $P(C_{\ell}|\sigma_{\ell})$,
with appropriate binning schemes for each of $\ct$, $\cx$ and $\ce$.
Again, producing a similar sample with the direct inverse Wishart
sampler is not possible, as discussed above.

\subsection{Marginal Blackwell-Rao estimators}
\label{sec:blackwell-rao}

Next, we consider Blackwell-Rao estimators for marginal and
conditional distributions, and focus for now on the factorization of
$P(C_{\ell}|\sigma_{\ell})$ given in equation \ref{eq:factorization}.
However, we note that any one of the distributions in Table
\ref{tab:distributions} gives rise to a separate estimator.

Recall first the derivation of the Blackwell-Rao estimator \citep{chu:2005},
\begin{align}
P(C_{\ell}|\mathbf{d}) &= \int P(C_{\ell}, \mathbf{s}|\mathbf{d})
\,d\mathbf{s}\notag\\ 
&= \int P(C_{\ell}|\mathbf{s},\mathbf{d})
P(\mathbf{s}|\mathbf{d})\,d\mathbf{s} \notag\\ 
&= \int P(C_{\ell}|\sigma_{\ell})
P(\sigma_{\ell}|\mathbf{d})\,D\sigma_{\ell}\notag\\ &
\approx \frac{1}{N_{\textrm{G}}}\sum_{i=1}^{N_{\textrm{G}}}
P(C_{\ell}|\sigma_{\ell}^{i}), 
\label{eq:br}
\end{align}
Thus, the full Blackwell-Rao estimator for $P(C_{\ell}|\mathbf{d})$ is
nothing but the sum (or average) of $P(C_{\ell}|\sigma_{\ell})$ over
Gibbs samples, $\sigma_{\ell}$.

\begin{figure}
\mbox{\epsfig{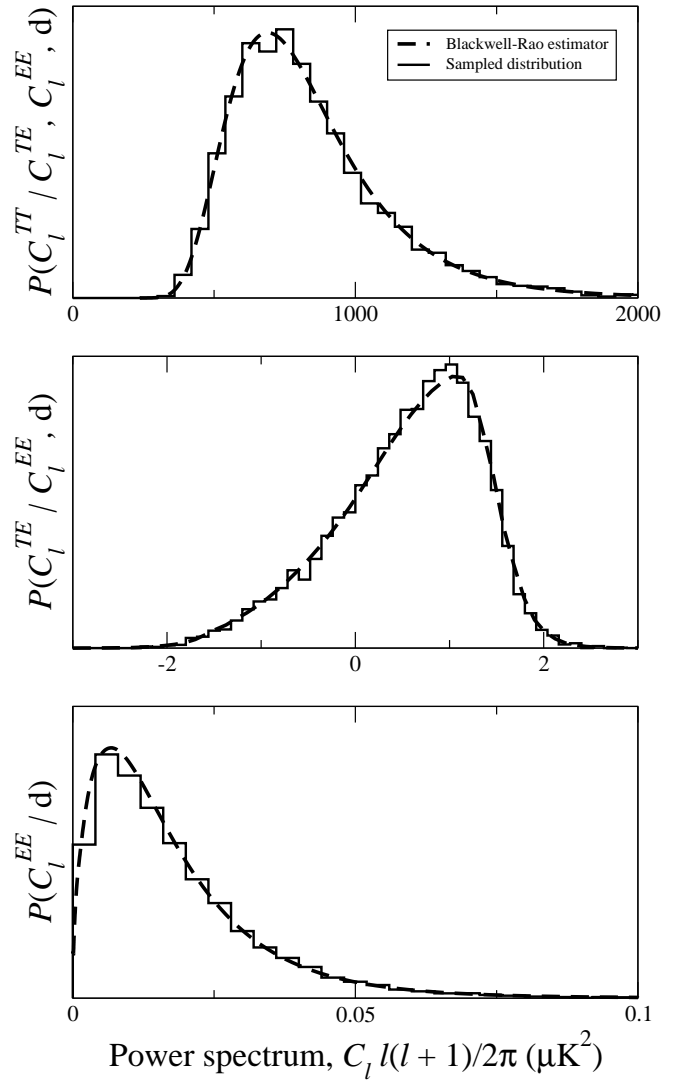}}
\caption{Comparison of three Blackwell-Rao estimators and simple
  histograms computed from a low-resolution CMB simulation.}
\label{fig:br}
\end{figure}

This estimator, as written here, has notoriously poor convergence
properties as the dimension of the parameter volume increases
\citep{chu:2005}: It requires an exponential number of samples in
order to converge. The reason is simply that the volume of a single
Gibbs sample is given by cosmic variance alone, whereas the volume of
the full distribution is determined also by noise and sky cut.
Therefore, even if the width of $P(C_{\ell}|\sigma_{\ell})$ for a
single multipole is as much as, say, 90\% of the width of
$P(C_{\ell}|\mathbf{d})$, the relative volume fraction in
$\ell_{\textrm{max}}$ dimensions is $f=0.9^{\textrm{lmax}}$. For
$\ell_{\textrm{max}} = 30$, this number is $f=0.04$, and for
$\ell_{\textrm{max}}=100$ it is $f=2.65\cdot10^{-5}$.  Clearly, a
brute-force Blackwell-Rao approach will not work for high-dimensional
spaces unless the volume ratio per multipole is unrealistically large.
However, the same problem does not affect the marginal distributions
described above, because the number of dimensions is low, and
typically just one, by construction.

Let us consider the Blackwell-Rao estimator for $P(\ce|\mathbf{d})$
for a single multipole, $\ell$. First, marginalization over other
multipoles consists, as usual for MCMC algorithms, simply of
disregarding the sample values of all other $\ell$'s. Second,
marginalization over $\ct$ and $\cx$ is done using the expressions
derived in Section \ref{sec:distributions},
\begin{align}
  P(\ce|\mathbf{d}) &= \iint P(\ct,\cx,\ce|\mathbf{d}) d\ct d\cx \notag\\
  & \approx \iint \sum_i P(\ct,\cx,\ce|\sel^i) d\ct d\cx \notag\\
  & = \sum_i \iint P(\ct,\cx,\ce|\sel^i) d\ct d\cx \notag\\
  & = \sum_i P(\ce|\sel^i) \notag\\
  & = \sum_i \frac{(\sigma_{\ell}^{i,EE})^{\frac{2\ell-3}{2}}}{(C_{\ell}^{EE})^{\frac{2\ell-1}{2}}}
e^{-\frac{2\ell+1}{2} \frac{\sigma_{\ell}^{i,EE}}{C_{\ell}^{EE}}}
\end{align}

Similarly, the Blackwell-Rao estimator for $P(\cx|\ce, \mathbf{d})$
reads 
\begin{align}
  &P(\cx|\ce, \mathbf{d}) =
  \frac{P(\cx,\ce|\mathbf{d})}{P(\ce|\mathbf{d})} \label{eq:TE_conditional}\\
  & \propto P(\cx,\ce|\mathbf{d}) \notag\\
  & = \sum_i P(\ct, \ce|\sel^i) \notag\\
  & \propto \sum_i \frac{(\ce)^{-1}|\sel^i|^{\frac{2\ell-2}{2}}
e^{-\frac{2\ell+1}{2}\frac{\sigma_{\ell}^{i,EE}}{\ce}}}
{\left(\sigma_{\ell}^{i,EE}\frac{(\cx)^2}{\ce}-2\sigma_{\ell}^{i,TE}
\cx+\sigma_{\ell}^{i,TT}\ce\right)^{\frac{2\ell-1}{2}}}
\end{align}
First, note that because $\ce$ is a constant in this expression,
$P(\ce|\mathbf{d})$ is also a constant, and can be disregarded after
equation \ref{eq:TE_conditional}. Second, we emphasize that it is
crucial to use the full joint expression for $P(\cx, \ct|\mathbf{d})$
in this estimator, and not the naive ``alternative'' $P(\cx|\ce,
\mathbf{d}) \approx \sum_i P(\cx|\ce,\sel^i)$; the latter approach
would require the underlying Gibbs samples, $\sel^i$, to be drawn
conditionally on $\ce$ in original Gibbs analysis, and this is usually
not the case.

Finally, we consider the expression for $P(\ct|\cx,\ce,\mathbf{d})$
for a single multipole. However, there is little simplification
involved in this expression, as it simply reads
\begin{align}
P(\ct|\ct, \ce, \mathbf{d}) &\propto P(\ct,\cx,\ce|\mathbf{d}) \\
& \approx \sum_i P(\ct,\cx,\ce|\sel^i) \notag\\
& = \sum_i \frac{|\sigma^i_{\ell}|^{2\ell-2}}{|C_{\ell }|^{\frac{2\ell+1}{2}}}
e^{-\frac{2\ell+1}{2}\textrm{tr}(\sigma^i_{\ell}C^{-1}_{\ell})},\notag
\end{align}
where $\sel$ and $C_{\ell}$ indicate 2-dimensional $\{T,E\}$ matrices.

As a simple illustration, we plot the three Blackwell-Rao estimators
given above for $\ell=10$ in Figure \ref{fig:br}, as computed from a
low-resolution simulation ($N_{\textrm{side}}=16$,
$\ell_{\textrm{max}}=47$, Gaussian beam of $10^{\circ}$ FWHM, and
white noise of $\sigma^{T} = 1\mu\textrm{K}$ and $\sigma^{Q,U} =
0.5\mu\textrm{K}$ for temperature and polarization, respectively)
drawn from a standard $\Lambda$CDM spectrum. As expected, the
agreement with direct histograms is excellent, but the smoothness and
faster convergence of the Blackwell-Rao estimator make it the
preferred choice for most applications.

While these expressions are useful in their own right, for example for
plotting marginal or joint power spectra and corresponding confidence
regions from a set of Gibbs samples (say, by maximizing and/or
integrating the Blackwell-Rao estimator for each $\ell$ individually),
their main application lies in providing robust estimates of
$P(\ct,\cx,\ce|\mathbf{d})$ in terms of univariates. This may open up
for some very interesting possibilities for stabilizing the
exponential behaviour of the full multivariate estimator. This topic
will be explored further in a future publication.

\section{Conclusions}
\label{sec:conclusions}

We have derived computationally convenient expressions for all
marginals of $P(C_{\ell}|\sigma_{\ell})$. These expressions may be
useful to any CMB analysis method that considers both the CMB sky
signal $\mathbf{s}$ and the power spectrum $C_{\ell}$ as unknown
parameters. One prominent example is the CMB Gibbs sampler.

We have also presented two specific applications of these expressions.
First, we demonstrated a new sampling algorithm for
$P(C_{\ell}|\sigma_{\ell})$ that supports different binning schemes
for each polarization component. This is useful because most
experiments have very different signal-to-noise ratio to temperature
and polarization.

Second, we have provided explicit expressions for the Blackwell-Rao
estimators for $P(\ct|\cx,\ce,\mathbf{d})$, $P(\cx|\ce, \mathbf{d})$
and $P(\ce|\mathbf{d})$. Together, these three can be used to map out
the joint distribution $P(C_{\ell}|\mathbf{d})$ for a single multipole
in terms of univariate distributions alone. Further, we note that any
of the distributions listed in Table \ref{tab:distributions} give rise
a separate, and potentially useful, Blackwell-Rao estimator.

\begin{acknowledgements}
  We thank Jeff Jewell, Greg Huey, Kris G{\'o}rski and Graca Rocha for
  useful and interesting discussions. We acknowledge use of the
  HEALPix\footnote{http://healpix.jpl.nasa.gov} software
  \citep{gorski:2005} and analysis package for deriving the results in
  this paper. We acknowledge use of the Legacy Archive for Microwave
  Background Data Analysis (LAMBDA). The authors acknowledge financial
  support from the Research Council of Norway.
\end{acknowledgements}

\end{document}